\documentclass[journal,twocolumn]{IEEEtran-v17}
\pdfoutput=1

\newlength{\figwidth}
\usepackage{color}
\definecolor{links}{rgb}{0.7,0,0}   
\definecolor{urls}{rgb}{0,0,0.8}    
\definecolor{cites}{rgb}{0,0,0.8}   

\usepackage[colorlinks,hyperindex,linkcolor=links,citecolor=cites,urlcolor=urls]{hyperref} 
 
\usepackage[nosort]{cite}        
\usepackage{url} 
\usepackage[intlimits]{amsmath}
\usepackage{bbm}
\usepackage{graphicx}
\usepackage{paralist}
\usepackage{fancyref}
\usepackage[stretch=16,shrink=16,step=4]{microtype}
\usepackage{vmr-symbols-vecbold}
\usepackage{standard-macros}
\usepackage{steinmetz}
\usepackage{dsfont}

\makeatletter
\def\@IEEEinterspaceratioM{0.265}
\def\@IEEEinterspaceMINratioM{0.1651}
\def\@IEEEinterspaceMAXratioM{0.38}

\def\@IEEEinterspaceratioB{0.31}
\def\@IEEEinterspaceMINratioB{0.19}
\def\@IEEEinterspaceMAXratioB{0.38}
\@IEEEtunefonts
\makeatother
\safemath{\gammadist}{\mathrm{Gamma}}  
\safemath{\zpar}{t_{\snr,\snr_0}}
\safemath{\outpdf}{q_{\vecy}}  
\newcommand{\givenalt}{\ \bigl\lvert \ }
\newcommand{\indfun}{\mathbbmss{1}}
\begin{document}

\IEEEoverridecommandlockouts

\title{Capacity bounds for MIMO microwave backhaul links affected by  phase noise}
%
%
\author{Giuseppe~Durisi,~\IEEEmembership{Senior Member,~IEEE,}
        Alberto~Tarable,~\IEEEmembership{Member,~IEEE,}    
        Christian~Camarda,~\IEEEmembership{Student Member,~IEEE,}
        Rahul~Devassy, 
        Guido Montorsi,~\IEEEmembership{Senior Member,~IEEE} 
 \thanks{G. Durisi and R. Devassy are with the Department of Signals and Systems, Chalmers University of Technology, Gothenburg, Sweden (e-mail: \{durisi,devassy\}@chalmers.se).}
 \thanks{A. Tarable is with IEIIT-CNR, Turin, Italy (e-mail: alberto.tarable@ieiit.cnr.it).}
 \thanks{C. Camarda and G. Montorsi are with the Department of Electronics and Telecommunications, Politecnico di Torino, Turin, Italy (e-mail: (\{christian.camarda,guido.montorsi\}@polito.it)}
 \thanks{The material in this paper was presented in part at the Information Theory and Application Workshop (ITA), San Diego, CA, Feb. 2013.}
 \thanks{This work was partially supported by the Swedish Agency for Innovation Systems (VINNOVA), under the project P36604-1 MAGIC.}
}%

%
%
%
\maketitle

\begin{abstract}
We present bounds and a closed-form high-SNR expression for the capacity of multiple-antenna systems affected by Wiener phase noise.
Our results are developed for the scenario where a single oscillator drives all the radio-frequency circuitries at each transceiver (common oscillator setup),
the input signal is subject to a peak-power constraint, and the channel matrix is deterministic.
This scenario is relevant for line-of-sight multiple-antenna microwave backhaul links with sufficiently small antenna spacing at the transceivers.
For the  $2\times 2$ multiple-antenna case,  for a Wiener phase-noise process with standard deviation equal to $6^\circ$, and  at the medium/high SNR values at which microwave backhaul links  operate, the upper bound reported in the paper exhibits a $3\dB$ gap from a lower bound obtained using $64$-QAM.
Furthermore, in this SNR regime the closed-form high-SNR expression is shown to be accurate.
\end{abstract}

\section{Introduction} 
\label{sec:introduction}
A cost-effective solution to the problem of guaranteeing backhaul connectivity in mobile cellular networks is the use of point-to-point microwave links in the Q and E bands~\cite{coldrey12-09a,hansryd11-a}.
The current terrific rate of increase in mobile data traffic makes these microwave radio links a potential bottleneck in the deployment of high-throughput cellular networks.
This consideration has stimulated a large body of research aimed at the design of high-capacity backhaul links~\cite{mehrpouyan12-09a,chinnici11-06a,tarable13-06a,durisi13-06a}.
One design challenge is that the use of high-order constellations  to increase throughput ($512$ QAM has been recently demonstrated in commercial products) makes the overall system extremely sensitive to \emph{phase noise}, i.e., to phase and frequency instabilities in the radio-frequency (RF) oscillators used at the transmitter and the receiver.   

A fundamental way to characterize the impact of phase noise on the throughput of microwave backhaul links is to study their Shannon capacity.
Unfortunately, the capacity of the phase-noise channel is not known in closed form even for simple channel models, although capacity bounds and asymptotic results in the limiting regime of high SNR have been reported in the literature.
Lapidoth characterized the capacity  of the general class of stationary phase-noise channels (the widely used Wiener model~\cite{colavolpe12-09a} belongs to this class) in the high-SNR regime~\cite{lapidoth02-10a}.
Specifically, he showed that whenever the phase-noise process has finite differential-entropy rate, the high-SNR capacity is equal to half the capacity of an AWGN channel with the same SNR, plus a correction term that accounts for the memory in the phase-noise process.
This result has been recently extended to the \emph{waveform} phase-noise channel in~\cite{ghozlan13-07a,ghozlan13-03a}.
The high-SNR capacity of the block-memoryless phase-noise channel (a non-stationary channel) has been characterized in~\cite{nuriyev05-03a,durisi12-08a}.
§

For the non-asymptotic regime of finite SNR, only capacity bounds are available in the literature.
Katz and Shamai~\cite{katz04-10a} provided tight  upper and lower bounds on the capacity of the memoryless phase-noise channel.
They also established that the capacity-achieving distribution is discrete with an infinite number of mass points.
Some of the bounds reported in~\cite{katz04-10a} have been  extended to the block-memoryless phase-noise case in~\cite{nuriyev05-03a,durisi12-08a}.
For the Wiener phase-noise model, an upper bound on the rates achievable with PSK constellations has been proposed in~\cite{barbieri11-12a}.
Capacity lower bounds obtained by numerically computing the information rates achievable with various families of finite-cardinality independent and identically distributed (\iid) input processes (e.g., QAM, PSK, and APSK constellations) have been reported e.g., in~\cite{barbieri11-12a,colavolpe12-09a,barletta12-05a}.
The numerical evaluation of these bounds is often based on the algorithm for the computation of the information rates for finite-state channels proposed in~\cite{arnold06-08a} (see~\cite{barletta13-07a} for different approaches based, e.g., on particle filtering).
Although the effect of phase noise in the \emph{measurement} of MIMO channels has been extensively investigated in the literature (see, e.g.,~\cite{baum11-11a}), little is known about the impact of phase noise on the MIMO capacity.
In~\cite{durisi13-06a}, it was shown that different RF circuitries configurations (e.g., independent oscillators at each antenna as opposed to a single oscillator driving all antennas) yield different spatial multiplexing gains.
The impact of phase noise on the achievable rates of massive MIMO systems has been recently discussed in~\cite{bjornson13-07a,pitarokoilis12-10a}.
However, the observations reported in~\cite{bjornson13-07a,pitarokoilis12-10a} are based only on capacity lower bounds and are not conclusive.

\subsection*{Contributions} 
\label{sec:contributions}
We study the capacity of multiple-antenna systems affected by phase noise.
Specifically, we consider the scenario where a single oscillator drives all RF circuitries at each transceiver.
We present a non-asymptotic capacity upper bound for the case of Wiener phase noise and the practically relevant scenario when the transmit codewords are subject to a peak-power constraint, which is more stringent than the average-power constraint analyzed so far in the phase-noise literature.
This upper bound improves on the one recently reported in~\cite{durisi13-02a}, which was derived under the assumption of codewords subject to an average-power constraint.
When particularized to constant-modulus constellations and to single-antenna systems, our bound recovers the upper bound obtained in~\cite{barbieri11-12a}.
We compare our upper bound with lower bounds obtained by evaluating numerically the information rates achievable with QAM constellations.
For the case of a Wiener phase-noise process with standard deviation of the phase increments equal to $6^\circ$, the gap between our upper bound and the information rates achievable with 64-QAM is about $3\dB$ for medium/high SNR values. 

We also provide a capacity characterization in the high-SNR regime that is accurate up to a term that vanishes as SNR grow large.
This characterization yields a capacity approximation that turns out to be accurate already at moderate SNR values.

%


\section{System Model} 
\label{sec:system_model}

\subsection{The input-output relation} 
\label{sec:the_input_output_relation}


We consider the following $M\times M$ MIMO phase-noise channel with memory
\begin{align}\label{eq:io}
  \vecy_k=e^{j\theta_k}\matH \vecx_k + \vecw_k, \quad k=1,2,\dots, n.
\end{align}
Here, $\vecx_k$ denotes the $M$-dimensional input vector at discrete time $k$; $\matH$ is the MIMO channel matrix, which we assume deterministic, full-rank, and known to the transmitter and the receiver; $\{\theta_k\}$ is the phase-noise process; and $\{\vecw_k\}$ is the additive Gaussian noise, which we assume independent and identically distributed (\iid) circularly symmetric with zero mean and covariance matrix $\matI_M$, i.e., $\vecw_k\distas\jpg(\veczero,\matI_M)$.
The input-output relation~\eqref{eq:io} describes accurately line-of-sight (LoS) MIMO backhaul links in which the antenna spacing at the transceivers is sufficiently small for the RF circuitries at each antenna to be driven by the same oscillator~\cite{durisi13-06a}.
We elaborate further on the accuracy of the model~\eqref{eq:io} in~\fref{sec:line_of_sight_mimo}.

We assume that the phase-noise samples $\{\theta_k\}$ form a Wiener process~\cite{colavolpe12-09a}, i.e.,\footnote{See~\cite{ghozlan13-07a} for a discussion on the limitations of this model.}
\begin{align}\label{eq:wiener_model}
  \theta_{k+1}=\theta_k +\Delta_k
\end{align}
where $\{\Delta_k\}$ is a sequence of \iid zero-mean Gaussian random variables with variance $\sigma_{\Delta}^2$, i.e., $\Delta_k\distas\normal(0,\sigma_{\Delta}^2)$ and the sum in~\eqref{eq:wiener_model} is modulo $2\pi$.
The \iid assumption on $\{\Delta_k\}$ implies that $\{\theta_k\}$ is a Markov process.
Specifically, 
 \begin{align}
   f_{\theta_k \given \theta_{k-1},\dots,\theta_0}=f_{\theta_k \given \theta_{k-1} }=f_{\Delta}
 \end{align}
where
\begin{align}\label{eq:pdf_wiener_increment}
  f_{\Delta}(\delta)\define\sum_{l=-\infty}^{\infty} \frac{1}{\sqrt{2\pi \sigma^2_{\Delta}}} \exp\lefto(-\frac{(\delta-2\pi l)^2 }{2\sigma^2_{\Delta}}\right), \quad \delta \in [0,2\pi].
\end{align}
In words, $f_{\Delta}$ is the probability density function (pdf)  of the innovation $\Delta_k$ modulo~$2\pi$.

Under the additional assumption that the initial phase $\theta_0$ is uniformly distributed in the interval $[0,2\pi]$, i.e., $\theta_0\distas \mathcal{U}[0,2\pi]$, the process~$\{\theta_k\}$ is stationary.
%
Let $\Delta \distas f_{\Delta}$ (defined in~\eqref{eq:pdf_wiener_increment}). 
The differential entropy rate of a stationary Wiener process is then given by
\begin{align}
  \difent(\{\theta_k\}) = h(\Delta).
\end{align}
%

\subsection{LoS MIMO} 
\label{sec:line_of_sight_mimo}
%
%
The idea behind LoS MIMO is to achieve a full-rank channel matrix \matH over a LoS link by a careful placement of the antennas at the transceivers~\cite{driessen99-02a,gesbert02-12a,bohagen07-04a}.
Indeed, when the antenna spacing $d$  at the transmitter and the receiver satisfies
\begin{align}\label{eq:interantenna_spacing}
  d\approx \sqrt{\lambda R/M}
\end{align}
where $\lambda$ is the wavelength and $R$ denotes the distance between the transmitter and the receiver, the channel matrix $\matH$ can be made not only full-rank, but also \emph{unitary}~\cite{gesbert02-12a,bohagen07-04a}.

We next discuss some implications of~\eqref{eq:interantenna_spacing} on the design of microwave backhaul links.
Consider a microwave backhaul link operating in the E band at $80$~GHz.  
Assume that the transceivers are equipped with $2$ antennas each and are $500$~m apart.
According to~\eqref{eq:interantenna_spacing}, the antenna spacing that results in a unitary channel matrix is about $97$ cm, which is compatible with the assumption of using a single oscillator to drive the RF circuitries of both antennas.
In some cases, it may be convenient to locate the two antennas closer than what~\eqref{eq:interantenna_spacing} prescribes. 
Then, \matH ceases to be unitary, although it can still be made full rank~\cite{bohagen07-04a}.
For a microwave backhaul link operating at  $20$~GHz over a $3$~Km link, \eqref{eq:interantenna_spacing} results in an antenna spacing of about $3.8$~m, which calls for a distributed oscillator solution.

As pointed out in~\fref{sec:the_input_output_relation}, in this paper we will focus exclusively on the single-oscillator scenario.
We will consider both the case of \matH unitary (our results will be somewhat sharper under this assumption), and the more general case of \matH  full-rank but not necessarily unitary.
The distributed oscillator setup will not be analyzed. 
Preliminary results concerning this scenario can be found in~\cite{durisi13-06a,bjornson13-07a,pitarokoilis12-10a}.


\subsection{Peak-amplitude constraint} 
\label{sec:peak_amplitude_constraint}
%
The results currently available on the capacity of phase noise channels~\cite{lapidoth02-10a,katz04-10a,nuriyev05-03a,durisi12-08a,durisi13-02a} were derived under the assumption that each transmit codeword $(\vecx_1, \dots, \vecx_n)$ is subject to the average-power constraint~\cite[Eq.~(9.2)]{cover06-a}
\begin{align}\label{eq:avp}
  \frac{1}{n}\sum_{k=1}^{n} \vecnorm{\vecx_k}^2\leq \snr.
\end{align}

In practice, each codeword entry $\vecx_k$ must obey a given \emph{peak-power constraint} to avoid distortions due to nonlinearities and saturation effects at the high-power amplifier~\cite{wunder12-12a}.
To account for this and obtain capacity results that are more relevant in practice, in this paper we substitute~\eqref{eq:avp} with  the more stringent peak-power constraint
\begin{align}\label{eq:pap}
    \vecnorm{\vecx_k}^2\leq \snr,\quad k=1,\dots,n.
\end{align}
The peak-power constraint~\eqref{eq:pap} has been considered previously in the information-theoretic literature, but not in the contest of phase-noise channels.
Smith~\cite{smith71-a} proved that the capacity-achieving distribution of an AWGN channel subject to~\eqref{eq:pap} is discrete with a finite number of mass point (in contrast, the capacity-achieving distribution under~\eqref{eq:avp}~\cite[Ch.~9]{cover06-a} is Gaussian).
More recently, Lapidoth~\cite{lapidoth05-07a} characterized the high-SNR capacity of single-antenna stationary fading channels subject to~\eqref{eq:pap} in the setting where no \emph{a priori} channel-state information is available at the receiver. 


\subsection{Channel Capacity} 
\label{sec:channel_capacity}
We are interested in computing the capacity of the MIMO phase-noise channel~\eqref{eq:io}, which---under the peak-amplitude constraint~\eqref{eq:pap}---is given by
\begin{align}\label{eq:capacity}
  C(\snr)=\lim_{n\to\infty}\frac{1}{n}\sup I(\vecy^n;\vecx^n)
\end{align}
where $\vecx^n=(\vecx_1, \dots, \vecx_n)$ and, similarly, $\vecy^n=(\vecy_1, \dots, \vecy_n)$.
Here, the supremum is over all probability distributions on $\vecx^n$ that satisfy~\eqref{eq:pap} with probability one (\wpone).
In \fref{sec:the_unitary_case}, we analyze $C(\snr)$ for the case of \matH being unitary.
The general full-rank case will be discussed in~\fref{sec:the_non_unitary_case}.

\section{The Unitary Case} 
\label{sec:the_unitary_case}

\subsection{Capacity Upper Bound} 
\label{sec:capacity_upper_bound}
We next present an upper bound on $C(\snr)$ that  improves on the one reported in~\cite{durisi13-02a} for the average-power constrained case.
With some minor adjustments, the bound turns out to be tight in the high-SNR regime (see \fref{sec:asymptotic_behavior}).

Before presenting our upper bound, two observations are in order.
\begin{enumerate}[i)]
\item As $\matH$ is known to transmitter and receiver,~$C(\snr)$ depends on $\matH$ only through its singular values.
Since $\matH$ is unitary, all singular values are equal to one.
Hence, we can (and will) assume without loss of generality that $\matH=\matI_{M}$.
%
\item In the following proposition, we establish that the capacity-achieving input process $\{\vecx_k\}$
can be assumed isotropically distributed, a property that will be useful in our analysis.
\end{enumerate}
\begin{propo}\label{propo:isotropic_distribution}
The input process $\{\vecx_k\}$ that achieves the capacity of the channel~\eqref{eq:io} when~$\matH$ is  unitary can be assumed isotropically distributed.
Specifically, if $\{\vecx_k\}$ achieves $C(\snr)$ in~\eqref{eq:capacity}, then $\{\matU_k\vecx_k\}$, where the matrix-valued random process $\{\matU_k\}$ is \iid and each $\matU_k$ is uniformly distributed on the set of $M\times M$ unitary matrices, achieves $C(\snr)$ as well.
\end{propo}  
\begin{IEEEproof}
 The proof, which exploits that $\matU_k\vecw_k\distas \vecw_k$, follows the same steps as the proof of~\cite[Prop.~7]{moser09-06a}. 
\end{IEEEproof}

Our upper bound on $C(\snr)$ is constructed by extending to the MIMO case the method used in~\cite{lapidoth02-10a} to derive an asymptotic bound on the capacity of stationary single-antenna phase-noise channels.
We also use the approach proposed in~\cite{katz04-10a,durisi12-08a} to make the bound non-asymptotic, and some of the tools developed in~\cite{lapidoth06-02a} to account for the presence of the peak-power constraint~\eqref{eq:pap}.

For convenience, we introduce the following notation: 
for every $a>0$, we let
\begin{align} \label{eq:phi_definition}
  \phi_l(a)\define\phase{1+z_l/\sqrt{a}}
\end{align}
where 
the random variables $\{z_l\}_{l\in \integers}$ are \iid $\jpg(0,1)$-distributed and $\phase{x}$ denotes the phase of the complex number $x$.
Roughly speaking, $\phi_l(a)$ is the noise level in the estimation of the phase-noise sample $\theta_l$ from the channel output $\vecy_l$ given that the input vector $\vecx_l$ is known and $\vecnorm{\vecx_l}^2=a$.

\begin{thm}\label{thm:capacity_upper_bound}
The capacity of the channel~\eqref{eq:io} under the peak-power constraint~\eqref{eq:pap} can be upper-bounded as
$C(\snr)\leq U(\snr)$, where\footnote{Throughout the paper, $\log$ stands for the natural logarithm.}
\begin{IEEEeqnarray}{rCL}\label{eq:upper_bound}
  U(\snr)&\define&  \min_{\alpha>0}\biggl\{ \alpha\log\frac{\snr+M}{\alpha} +
	d_{\alpha}+\log(2\pi) \notag\\
  &&+\max_{0\leq \xi \leq \sqrt{\snr}} g_{\alpha}(\xi,\snr) \biggr\}.
\end{IEEEeqnarray}
Here,
\begin{IEEEeqnarray}{rCL}
  \IEEEeqnarraymulticol{3}{l}{g_{\alpha}(\xi,\snr)\define (M-\alpha)\Ex{}{\log\lefto(\abs{\xi+z_1}^2 +\sum_{j=2}^M \abs{z_j}^2\right) }}
 	  \notag\\
    &&+\alpha \frac{\xi^2+M}{\snr+M}-\difent(\abs{\xi+z_0}^2)\notag\\ 
  &&-\difent\lefto(\theta_0+\phi_0(\xi^2) \givenalt \left\{ \theta_l +\phi_{l}(\snr) \right\}_{l=-\infty}^{-1},\abs{\xi+z_0}\right) \label{eq:g_definition}
\end{IEEEeqnarray}
where  $\{z_l\}$ are \iid  $\jpg(0,1)$-distributed random variables and
\begin{align}\label{eq:constant_ub}
	d_{\alpha}\define\log\frac{\Gamma(\alpha)}{\Gamma(M)}-M+1
\end{align}
with $\Gamma(\cdot)$ standing for the Gamma function.
\end{thm}
\begin{IEEEproof}
Because of Proposition~\ref{propo:isotropic_distribution}, we can restrict ourselves to  isotropically distributed input processes.
Specifically, we consider $\{\vecx_k\}$ of the form $\{\vecx_k=s_k\vecv_k\}$, where $s_k=\vecnorm{\vecx_k}$ and $\vecv_k=\vecx_k/s_k$, with $\vecv_k$ uniformly distributed on the unit sphere in $\complexset^M$ and independent of $s_k$. 
We start by using chain rule as follows
\begin{IEEEeqnarray}{rCL}\label{eq:chain_rule_ub}
  I(\vecy^n;\vecx^n) &=& \sum_{k=1}^{n} I(\vecy_k;\vecx^n \given \vecy^{k-1}).
\end{IEEEeqnarray}
By proceeding similarly to~\cite[Eq.~(10)]{durisi13-02a}, but accounting for the peak-power constraint,\footnote{Recall the the upper bound developed in~\cite{durisi13-02a} holds for the average-power constraint case.}  we next upper-bound each term on the right-hand side (RHS) of~\eqref{eq:chain_rule_ub}. 
We first note that
\begin{IEEEeqnarray}{rCL}
  I( \vecy_k;\vecx^n \given \vecy^{k-1})
   &=& 
   h(\vecy_k \given \vecy^{k-1}) - h(\vecy_k \given \vecy^{k-1}, \vecx^n) \IEEEeqnarraynumspace\\
    &{\leq}& h(\vecy_k) - h(\vecy_k \given \vecy^{k-1}, \vecx^n)\label{eq:easy_step_in_upper_bound_step_a} \IEEEeqnarraynumspace\\
      &{=}& h(\vecy_k) - h(\vecy_k \given \vecy^{k-1}, \vecx^{k-1}, \vecx_k). \IEEEeqnarraynumspace \label{eq:easy_step_in_upper_bound}
    %
\end{IEEEeqnarray}
Here, in~\eqref{eq:easy_step_in_upper_bound_step_a} we used that conditioning reduces entropy, and in~\eqref{eq:easy_step_in_upper_bound} that $\vecy_k$ and $(\vecx_{k+1},\dots,\vecx_n)$ are conditionally independent given $(\vecy^{k-1},\vecx^k)$.
We next focus on the conditional differential entropy (the second term) on the RHS of~\eqref{eq:easy_step_in_upper_bound}.
Intuitively, the past inputs $\vecx^{k-1}$ and the past outputs $\vecy^{k-1}$ can be used to obtain noisy estimates of the past phase-noise samples $\{\theta_l\}_{l=1}^{k-1}$.
These estimates help us to guess the value of the current phase-noise sample $\theta_k$. 
We next use this intuition to obtain a lower bound on $h(\vecy_k \given \vecy^{k-1}, \vecx^{k-1}, \vecx_k)$, and, hence, an upper bound on $I(\vecy_k; \vecx^n \given \vecy^{k-1})$ in~\eqref{eq:chain_rule_ub}.
For each pair  $(\vecy_l,\vecx_l)$, $l=1,\dots,k-1$, we compute the phase  of the projection of $\vecy_l$ onto $\vecx_l$.
This projection is distributed as $\theta_l+\phi_l(s_l^2)$.
Since $\vecy_k$ and $\vecy^{k-1}$ are conditionally independent given both $\vecx^k$ and $\{\theta_l+\phi_l(s_l^2)\}_{l=1}^{k-1}$, we obtain
\begin{IEEEeqnarray}{rCL}
  \IEEEeqnarraymulticol{3}{l}{ h(\vecy_k \given \vecy^{k-1}, \vecx^{k-1}, \vecx_k)}\\
  \quad&=& h\lefto(\vecy_k \givenalt \bigl\{\theta_l + \phi_l(s_l^2) \bigr\}_{l=1}^{k-1}, \vecx^{k-1}, \vecx_k\right) \\
  \quad&\geq & h\lefto(\vecy_k \givenalt \bigl\{\theta_l+ \phi_l(\snr) \bigr\}_{l=1}^{k-1}, \vecx_k\right).\label{eq:bound_on_diff_ent_exploiting_peak}
\end{IEEEeqnarray}
In the last step, we used that the best noisy estimate of the past phase-noise samples $\{\theta_l\}_{l=1}^{k-1}$ is achieved by transmitting inputs at peak power, i.e., $s_l^2=\snr$, $l=1,\dots,k-1$.
Substituting~\eqref{eq:bound_on_diff_ent_exploiting_peak} into~\eqref{eq:easy_step_in_upper_bound}, we obtain
\begin{IEEEeqnarray}{rCL}
  I(\vecy_k; \vecx^n \given \vecy^{k-1})
   &\leq& \difent(\vecy_k) \notag \\
   &&-\ \difent\lefto(\vecy_k \givenalt \bigl\{\theta_l+\phi_l(\snr) \bigr\}_{l=1}^{k-1},\vecx_k\right) \\
   &=& I\lefto(\vecy_k;\bigl\{\theta_l+\phi_l(\snr) \bigr\}_{l=1}^{k-1},\vecx_k\right)\\
   &=& I\lefto(\vecy_0;\bigl\{\theta_l+\phi_l(\snr) \bigr\}_{l=-(k-1)}^{-1},\vecx_0\right)\label{eq:ub_split_into_two_terms_step_a}\\
   &\leq & I\lefto(\vecy_0;\bigl\{\theta_l+\phi_l(\snr) \bigr\}_{l=-\infty}^{-1},\vecx_0\right)\\
   &=& I(\vecy_0;\vecx_0) \notag\\
   &&+\ I\lefto(\vecy_0; \bigl\{\theta_l+\phi_l(\snr) \bigr\}_{l=-\infty}^{-1} \givenalt \vecx_0\right). \IEEEeqnarraynumspace\label{eq:ub_split_into_two_terms}
\end{IEEEeqnarray}
Here,~\eqref{eq:ub_split_into_two_terms_step_a} follows because $\{\theta_k\}$ is a stationary process.
Substituting~\eqref{eq:ub_split_into_two_terms} into~\eqref{eq:chain_rule_ub} and then~\eqref{eq:chain_rule_ub} into~\eqref{eq:capacity}, we obtain
\begin{multline}\label{eq:k_independent_upper_bound}
  C(\snr)\leq \sup \Bigl\{ I(\vecy_0;\vecx_0) \\
  + I\lefto(\vecy_0;\left\{\theta_l+\phi_l(\snr) \right\}_{l=-\infty}^{-1}\givenalt \vecx_0\right) \Bigr\}.
\end{multline}
The supremum in~\eqref{eq:k_independent_upper_bound} is over all probability distributions on $\vecx_0=s_0\vecv_0$ such that $s_0$ and $\vecv_0$ are independent, $\vecv_0$ is uniformly distributed on the unit sphere in $\complexset^M$, and $s^2_0\leq \snr$ \wpone.

We next upper-bound the first term on the RHS of~\eqref{eq:k_independent_upper_bound}, which corresponds to the mutual information of a memoryless phase-noise channel with uniform phase noise using a method similar to the one used in~\cite{durisi12-08a,katz04-10a}.
Specifically, we use the duality approach~\cite[Th.~5.1]{lapidoth03-10a} and choose an output probability distribution for which $\vecy_0$ is isotropically distributed and $r=\vecnorm{\vecy_0}^2$ follows a Gamma distribution with parameters $\alpha$ to be optimized later and $\beta\define (\snr +M)/\alpha$.
To summarize the probability density function (pdf) of $r$ is given by
\begin{IEEEeqnarray}{rCL}\label{eq:output_distribution_avp}
  q_r(r)=\frac{r^{\alpha-1}e^{-r/\beta}}{\beta^{\alpha}\Gamma(\alpha)}.
\end{IEEEeqnarray}
This output distribution is optimal at high SNR (i.e., it achieves capacity up to a term that vanishes as SNR grows large) for the average-power constraint case~\cite{durisi13-02a}.
However, it is not optimal for the peak-power constraint case, as we shall discuss in the Appendix.
Nevertheless, it leads to a bound that is accurate for medium SNR values (see \fref{sec:simulation_results}).

Using~\eqref{eq:output_distribution_avp}, we upper-bound $I(\vecx_0;\vecy_0)$ as follows~(see~\cite{durisi13-02a}): 
\begin{IEEEeqnarray}{rCL}\label{eq:ub_bound_on_the_memoryless_part}
  I(\vecy_0;\vecx_0) &\leq&\alpha \log\frac{\snr +M}{\alpha}+d_{\alpha} \notag\\
        &&+ (M-\alpha)\Ex{}{\log\lefto(\abs{s_0+z_1}^2 + \sum_{j=2}^M\abs{z_j}^2\right)} \notag\\
        &&- \difent\lefto(\abs{s_0+z_0}^2 \givenalt s_0\right)+ \alpha \frac{\Ex{}{s_0^2}+M}{\snr +M}.
\end{IEEEeqnarray}
Here, $d_{\alpha}$ is the constant defined in~\eqref{eq:constant_ub} and $z_0,z_1,\dots,z_M$ are \iid $\jpg(0,1)$-distributed random variables. 

The second term on the RHS of~\eqref{eq:ub_split_into_two_terms} can be evaluated as follows:
\begin{IEEEeqnarray}{rCL}
  \IEEEeqnarraymulticol{3}{l}{I\lefto(\vecy_0; \bigl\{\theta_l+\phi_l(\snr) \bigr\}_{l=-\infty}^{-1} \givenalt \vecx_0\right)}\notag\\
  %
  &=&
 I\lefto(e^{j\theta_0}s_0+z_0;\bigl\{\theta_l+\phi_l(\snr) \bigr\}_{l=-\infty}^{-1} \given s_0\right)\label{eq:ub_bound_on_the_memory_part_step_a}\\
  &=& I\lefto(e^{j\theta_0}(s_0+z_0);\bigl\{\theta_l+\phi_l(\snr) \bigr\}_{l=-\infty}^{-1}\given s_0\right)\label{eq:ub_bound_on_the_memory_part_step_b}\\
  &=& I\lefto(\abs{s_0+z_0},\theta_0+\phi_0(s^2_0);\bigl\{\theta_l+\phi_l(\snr) \bigr\}_{l=-\infty}^{-1}\given s_0\right) \label{eq:ub_bound_on_the_memory_part_current_noise}\\
  &=& I\lefto(\theta_0+\phi_0(s^2_0);\bigl\{\theta_l+\phi_l(\snr) \bigr\}_{l=-\infty}^{-1} \givenalt \abs{s_0+z_0},s_0\right)\label{eq:ub_bound_on_the_memory_part_step_c}\\
  &=&\difent\lefto(\theta_0+\phi_0(s^2_0) \givenalt \abs{s_0+z_0},s_0\right) \notag\\
  &&- \difent\lefto(\theta_0+\phi_0(s^2_0) \givenalt \bigl\{\theta_l+\phi_l(\snr) \bigr\}_{l=-\infty}^{-1}, \abs{s_0+z_0},s_0\right) \\
  &=& \log(2\pi) \notag\\ 
  &&- \difent\lefto(\theta_0+\phi_0(s^2_0) \givenalt \bigl\{\theta_l+\phi_l(\snr) \bigr\}_{l=-\infty}^{-1}, \abs{s_0+z_0},s_0\right).\IEEEeqnarraynumspace\label{eq:ub_bound_on_the_memory_part_step_d}
  %
\end{IEEEeqnarray}
Here,~\eqref{eq:ub_bound_on_the_memory_part_step_a} follows because $\herm{\vecv_0}\vecy_0\distas e^{j\theta_0}s_0+z_0$ is a sufficient statistics for $\{\theta_l+\phi_l(\snr)\}_{l=-\infty}^{-1}$; \eqref{eq:ub_bound_on_the_memory_part_step_b}~follows because $z_0$ is circularly symmetric;~\eqref{eq:ub_bound_on_the_memory_part_step_c} holds because $\abs{s_0+z_0}$ and $\{\theta_l+\phi_l(\snr)\}_{l=-\infty}^{-1}$ are independent; finally,~\eqref{eq:ub_bound_on_the_memory_part_step_d} holds because $\theta_0\distas\setU[0,2\pi]$.

We  substitute~\eqref{eq:ub_bound_on_the_memoryless_part} and~\eqref{eq:ub_bound_on_the_memory_part_step_d} into~\eqref{eq:k_independent_upper_bound}, upper-bound the supremum over all probability distributions on $s_0$ satisfying $s_0\leq \sqrt{\snr}$ \wpone with the supremum over all deterministic $\xi \in [0,\sqrt{\snr}]$, and tighten the resulting bound by minimizing it over the optimization parameter $\alpha>0$.
This concludes the proof.
\end{IEEEproof}

\paragraph*{Remarks} 
\label{par:remarks_}
The coarser upper bound provided in~\cite[Th.~2]{durisi13-02a} can be obtained from $U(\snr)$ in~\eqref{eq:upper_bound} by assuming perfect knowledge of the past phase-noise samples.
This results in the following cruder lower bound on $\difent(\vecy_k\given \vecy^{k-1},\vecx^{k-1},\vecx_k)$ (cf.,~\eqref{eq:bound_on_diff_ent_exploiting_peak})
\begin{IEEEeqnarray}{rCL}
  \difent(\vecy_k\given \vecy^{k-1},\vecx^{k-1},\vecx_k) &\geq& \difent(\vecy_k \given \theta^{k-1},\vecx_k)\\
  &=& \difent(\vecy_k \given \theta_{k-1},\vecx_k).
\end{IEEEeqnarray}
An even coarser bound can be obtained by assuming perfect knowledge of the additive noise $\phi_0(s^2_0)$ affecting the current phase-noise sample (see~\eqref{eq:ub_bound_on_the_memory_part_current_noise}--\eqref{eq:ub_bound_on_the_memory_part_step_d}).
This results in the following simple capacity upper bound
\begin{IEEEeqnarray}{rCL}\label{eq:memoryless_plus_correction}
  C(\snr)\leq  \sup \left\{I(\vecy_0;\vecx_0) \right\} + \log(2\pi)- \difent(\Delta)
\end{IEEEeqnarray}
where $\Delta$ is distributed as in~\eqref{eq:pdf_wiener_increment}.
The inequality~\eqref{eq:memoryless_plus_correction} can be interpreted as follows: the capacity of a Wiener phase-noise channel is upper-bounded by the capacity of a memoryless phase-noise channel with uniform phase noise, plus a correction term that accounts for the memory in the channel and does not depend on the SNR~\snr.

If we now specialize~\eqref{eq:memoryless_plus_correction} to single-antenna systems and we add the additional constraint that $\abs{s}=\sqrt{\snr}$ \wpone (which holds, for example, if a PSK constellation is used), the first term on the RHS of~\eqref{eq:memoryless_plus_correction} vanishes and we recover the upper bound previously reported in~\cite[Th.~2]{barbieri11-12a}.

The last term in~\eqref{eq:g_definition} can be computed by using a slightly modified version of the algorithm described in~\cite{arnold06-08a}.


\subsection{Asymptotic Behavior} 
\label{sec:asymptotic_behavior}
In Theorem~\ref{thm:large_snr} below, we present an asymptotic characterization of $C(\snr)$ that generalizes to the MIMO case and to the case of peak-power-constrained inputs the asymptotic characterization reported in~\cite{lapidoth02-10a} for the single-antenna case and average-power-constrained inputs.
\begin{thm}\label{thm:large_snr}
In the high-SNR regime, the capacity of the Wiener phase-noise channel~\eqref{eq:io} behaves as
\begin{IEEEeqnarray}{rCL}
C(\snr)&=& \left(M-\frac{1}{2}\right)\log\snr - \log\lefto(M-\frac{1}{2}\right) -\log \Gamma(M)\notag\\
&&+\frac{1}{2}\log \pi-\left(M-\frac{1}{2}\right)-\difent(\Delta)+\landauo(1)\label{eq:capacity_asymptotic}
\end{IEEEeqnarray}
where $\landauo(1)$ indicates a function of $\snr$ that vanishes in the limit $\snr\to\infty$.
\end{thm}
\begin{IEEEproof}
  The proof, which is rather technical, is relegated to the appendix.
\end{IEEEproof}

\subsection{Average power versus peak power} 
\label{sec:average_power_versus_peak_power}
By comparing the asymptotic capacity expansion provided in \fref{thm:large_snr} with the one reported in~\cite[Th.~3]{durisi13-02a} for the case of average-power-constrained input signals, we can assess the throughput loss at high SNR due to the presence of the more stringent peak-power constraint~\eqref{eq:pap}.
Specifically, let $C_{\text{ap}}(\snr)$ denote the capacity of the channel in~\eqref{eq:io} when the input signal is subject to~\eqref{eq:avp} instead of~\eqref{eq:pap}.
Furthermore, let $C(\snr)$ as in~\eqref{eq:capacity}.
Then
\begin{multline}
  \lim_{\snr\to\infty} \left\{C_{\text{ap}}(\snr)-C(\snr)\right\}= \log\Gamma\lefto(M-\frac{1}{2}\right)\\- \left(M-\frac{3}{2}\right)\log\frac{1}{M-1/2} +\left(M-\frac{1}{2}\right).
\end{multline}
For the single-antenna case (i.e., $M=1$) this asymptotic capacity loss is about $1$ bit/s/Hz.

%
%

\section{The Non-unitary Case} 
\label{sec:the_non_unitary_case}
As mentioned in Section \ref{sec:line_of_sight_mimo}, practical considerations may force the channel matrix $\matH$ to be non-unitary. In this section, we derive non-asymptotic upper and lower bounds on $\capacity(\snr)$ for the general case of full rank $\matH$, which are function of the capacity for the case of  unitary $\matH$.
This allows us to extend the results reported in \fref{sec:the_unitary_case} to the non-unitary case.


Let $\lambda\sub{min}$ and $\lambda\sub{max}$ denote the smallest and the largest eigenvalue of $\herm{\matH}\matH$, respectively.
The following theorem gives upper and lower bounds to $\capacity(\snr)$ for arbitrary full-rank matrix.

\begin{thm}\label{thm:capacity_bounds_nonunitary}
Let $C_{\rm unitary}(\snr)$ be the capacity of the channel in \eqref{eq:io} for the case of unitary $\matH$. 
The capacity for the case of an arbitrary full-rank matrix \matH with smallest and largest singular values given by $\sqrt{\lambda\sub{min}}$ and $\sqrt{\lambda\sub{max}}$, respectively, can be bounded as follows:  
\begin{align} \label{eq:bounds_on_arbitrary_H_capacity}
  C_{\rm unitary}(\lambda\sub{min}\snr ) \leq \capacity(\snr) \leq  C_{\rm unitary}(\lambda\sub{max}\snr ).
\end{align}

\end{thm}
\begin{IEEEproof}
Since $\matH$ has full rank and is known at both sides, precoding at the transmitter can be done in order to invert the channel. Precisely, set $\vecx_k=\matH^{-1}\tilde{\vecx}_k$, so that~\eqref{eq:io} becomes 
\begin{align}
  \vecy_k=e^{j\theta_k}\tilde{\vecx}_k + \vecw_k, \quad k=1,2,\dots, n.
\end{align}
The peak-power constraint \eqref{eq:pap} forces $\tilde{\vecx}_k$ within the hyperellipsoid
\begin{align}\label{eq:pap_nounitary}
  \herm{\tilde{\vecx}}_k \left( \matH\herm{\matH}\right)^{-1} \tilde{\vecx}_k\leq \snr , \quad k=1,\dots,n
\end{align}
\wpone.
By definition:
\begin{align}
  C_{\rm unitary}( \lambda\sub{max}\snr )    
  =\lim_{n\to\infty}\frac{1}{n}\sup I(\tilde{\vecx}^n;\vecy^n)
\end{align}
where the supremum is over all distributions on $\tilde{\vecx}^n$ that satisfy 
\begin{align}\label{eq:peak-power constraint_upper_bound}
  \vecnorm{\tilde{\vecx}_k}^2\leq \lambda\sub{max}\snr , \quad k=1,\dots,n, \quad \wpone.
\end{align}
The peak-power constraint in~\eqref{eq:peak-power constraint_upper_bound} is looser than~\eqref{eq:pap_nounitary}. 
Indeed,
\begin{align}\label{eq:norm_inequality}
  \vecnorm{\tilde{\vecx}_k}^2 /\lambda\sub{max} \leq \herm{\tilde{\vecx}}_k \left( \matH\herm{\matH}\right)^{-1} \tilde{\vecx}_k \leq \vecnorm{\tilde{\vecx}_k}^2 /\lambda\sub{min}. 
\end{align}
Hence,  if~\eqref{eq:pap_nounitary} holds, then~\eqref{eq:peak-power constraint_upper_bound} holds as well.
This implies that   
\begin{align}\label{eq:ub_capacity}
  C(\snr)\leq C_{\rm unitary}( \lambda\sub{max}\snr ).
\end{align}
In the same way, \eqref{eq:norm_inequality} and the definition of $C_{\rm unitary}(\snr )  $ allow us to conclude that
\begin{align}\label{eq:lb_capacity}
  C(\snr)\geq C_{\rm unitary}(\lambda\sub{min}\snr ).
\end{align}
\end{IEEEproof}

As a consequence of~\eqref{eq:bounds_on_arbitrary_H_capacity}, bounds for the case of unitary \matH can be transformed into bounds for the case of full-rank \matH at the cost of a power offset. 
Using the asymptotic expression for  $C_{\rm unitary}(\snr )$ reported in \eqref{eq:capacity_asymptotic}, we see that in the high-SNR regime the gap between the upper and lower bounds in~\eqref{eq:bounds_on_arbitrary_H_capacity}  is equal to
\begin{multline}\label{eq:upper_lower_gap}
   C_{\rm unitary}(\lambda\sub{max}\snr ) -  C_{\rm unitary}(\lambda\sub{min}\snr ) \\= \left(M-\frac{1}{2}\right)\log\frac{\lambda\sub{max}}{\lambda\sub{min}} +\landauo(1)
\end{multline}
which tends to a constant as SNR  increases.

Note that both the upper and the lower bound are obtained by neglecting the actual structure of \eqref{eq:pap_nounitary}. 
In order to take this structure into account, a non-isotropic distribution of $\tilde{\vecx}_k$, with power allocated according to a waterfilling strategy, may result in a tighter lower bound. 

\section{Numerical Results} 
\label{sec:simulation_results}

\begin{figure}[h]
  \centering
    \includegraphics[width=\figwidth]{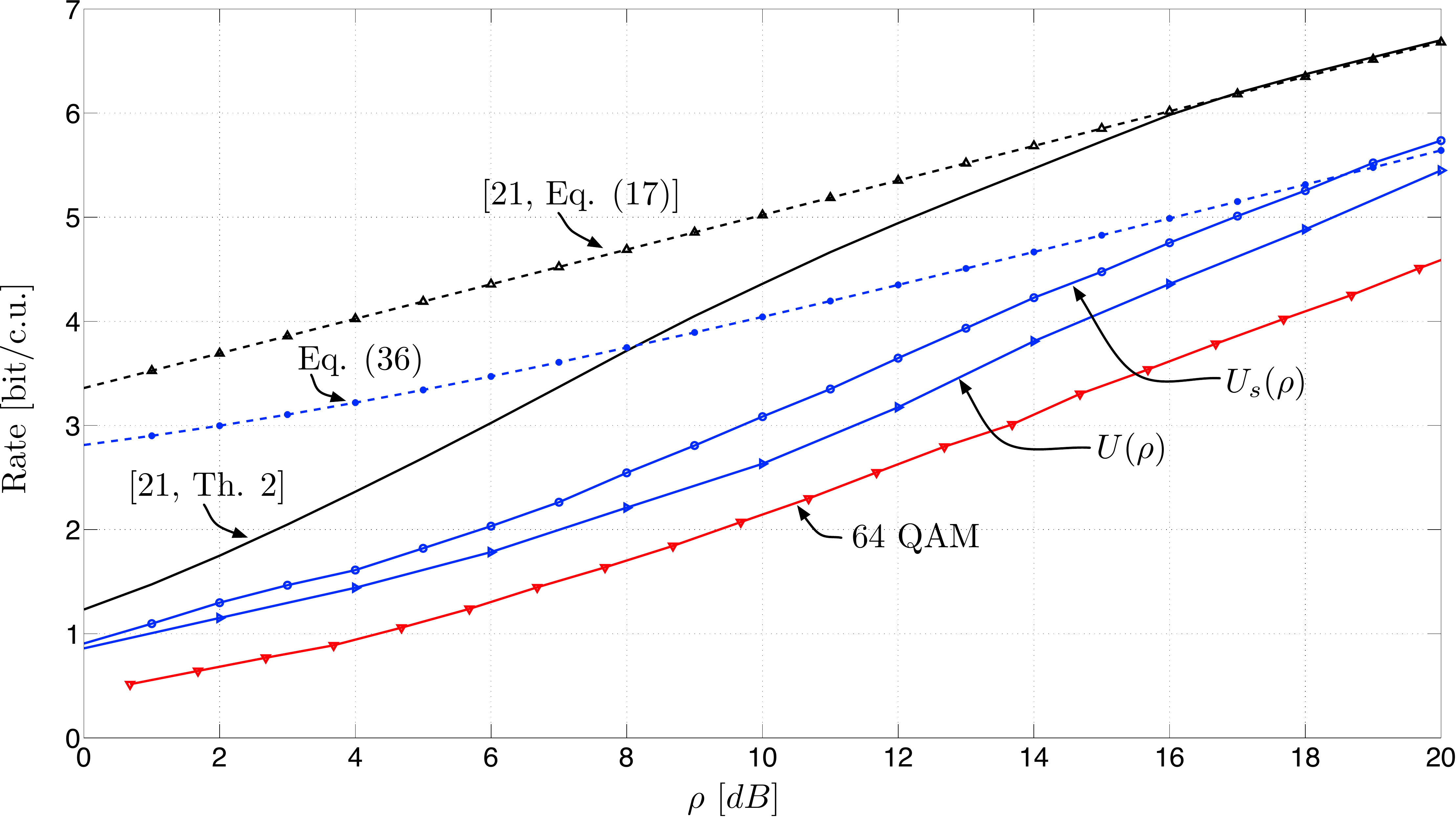}
  \caption{The upper bound $U(\snr)$ in~\eqref{eq:upper_bound}, its simplified version $U\sub{s}(\snr)$, the asymptotic capacity approximation~\eqref{eq:capacity_asymptotic}, the upper bound from~\cite[Th.~2]{durisi13-02a} and its asymptotic version~\cite[Eq.~(17)]{durisi13-02a}, and the rates achievable with  $64$-QAM. In the figure,~$\sigma_{\Delta}=6^{\circ}$.}
  \label{fig:6degSISO}
\end{figure}

In this section, we numerically compute the upper bound in~\eqref{eq:upper_bound} and compare it with the asymptotic expression in~\eqref{eq:capacity_asymptotic}, for a standard deviation of the phase-noise increments equal to $\sigma_{\Delta} = 6^{\circ}$, in the two cases $M = 1$ (single-antenna system) and $M=2$. 
In Fig.~\ref{fig:6degSISO}, the curves for the single-antenna case are displayed.
The bound approaches the asymptotic expression as SNR grows large, although it remains below it for all the SNR values considered.
In the figure, we also show the upper bound from~\cite[Th.~2]{durisi13-02a} and its asymptotic expansion~\cite[Eq.~(17)]{durisi13-02a}. 
Although these results were derived for an average-power constraint, they serve as upper bounds for the capacity under a peak-power constraint, since the latter is more stringent than the former. 
Finally, we also plot an upper bound that is obtained from~\eqref{eq:upper_bound} by substituting the conditional differential entropy 
\begin{IEEEeqnarray}{rCL}
  \difent\lefto(\theta_0+\phi_0(\xi^2) \givenalt \left\{ \theta_l +\phi_{l}(\snr) \right\}_{l=-\infty}^{-1},\abs{\xi+z_0}\right)
\end{IEEEeqnarray}
with
\begin{IEEEeqnarray}{rCL}
  \difent\lefto(\Delta+\phi_0(\xi^2) \givenalt \abs{\xi+z_0}\right).
\end{IEEEeqnarray}

This bound, which we refer to as $U\sub{s}(\snr)$ (where the letter ``s'' stands for ``simplified'') is much simpler to evaluate numerically than $U(\snr)$.
Furthermore, its computational complexity does not scale with the number of antennas (on the contrary, the computational complexity of $U(\snr)$ increases exponentially with the number of antennas).
 Unfortunately,$U\sub{s}(\snr)$ is less tight than $U(\snr)$  because
\begin{multline}
  \difent\lefto(\theta_0+\phi_0(\xi^2) \givenalt \left\{ \theta_l +\phi_{l}(\snr) \right\}_{l=-\infty}^{-1},\abs{\xi+z_0}\right) \\\geq \difent\lefto(\Delta+\phi_0(\xi^2) \givenalt \abs{\xi+z_0}\right).
\end{multline}

The newly derived bounds improve on the previous ones by 6-7 dB at moderate and high SNR values, in accordance with what reported in Subsection~\ref{sec:average_power_versus_peak_power}. 
Finally, the numerically computed mutual information for the case of 64-QAM is also shown.\footnote{Specifically, we use the algorithm for the computation of the information rates for finite-state channels proposed~\cite{arnold06-08a}. 
We choose 200 levels for the discretization of the phase-noise process, and average over a block of 2000 channel uses.} A gap ranging from $2\dB$ to $3\dB$ is observed between~\eqref{eq:upper_bound} and the 64-QAM curve, depending on the SNR.
From the plot, we see that the asymptotic capacity expression, which, differently from both upper and lower bounds, is trivial to compute, accurately describes the behavior of the capacity for SNR larger than $16\dB$.

\begin{figure}[h]
  \centering
    \includegraphics[width=\figwidth]{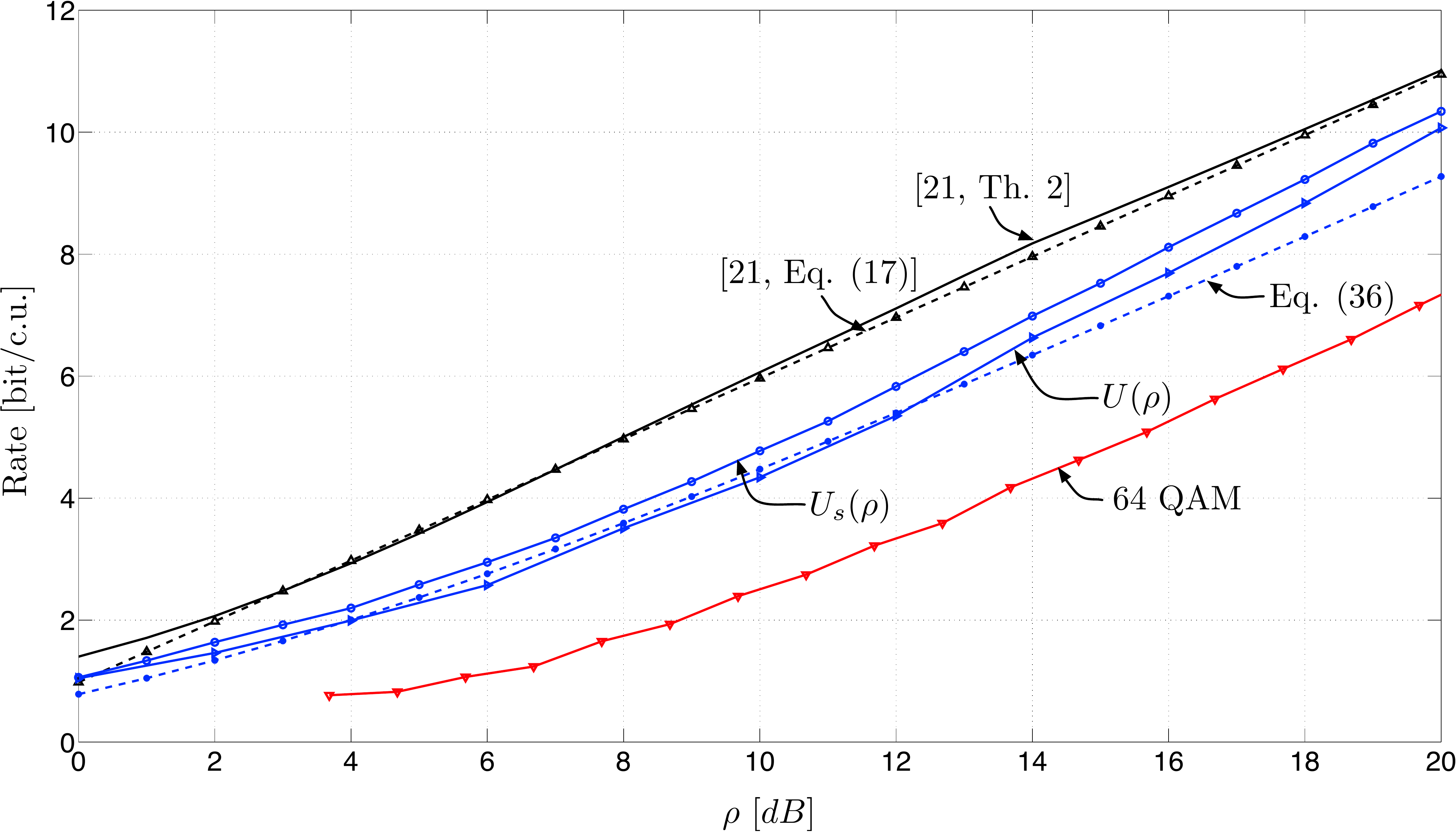}
  \caption{The upper bound $U(\snr)$ in~\eqref{eq:upper_bound}, its simplified version $U\sub{s}(\snr)$, the asymptotic capacity approximation~\eqref{eq:capacity_asymptotic}, the upper bound from~\cite[Th.~2]{durisi13-02a} and its asymptotic version~\cite[Eq.~(17)]{durisi13-02a}, and the rates achievable with  $64$-QAM. In the figure,~$\sigma_{\Delta}=6^{\circ}$.}
  \label{fig:6degMIMO}
\end{figure}

In Fig.~\ref{fig:6degMIMO}, the curves for the case $M=2$ are shown. As in Fig.~\ref{fig:6degSISO}, we also depict the upper bound from~\cite[Th.~2]{durisi13-02a} together with its asymptotic version~\cite[Eq.~(17)]{durisi13-02a}, the simplified upper bound $U_s(\snr)$, and the mutual information achieved by 64-QAM. 
The newly derived bounds improve on the previous ones by about $3\dB$  in the high-SNR region.  For the MIMO case, the gap between~\eqref{eq:upper_bound} and the QAM curve is about $3.5\dB$  in the high-SNR region and larger for smaller SNR values.
In this case, the asymptotic capacity expression seems to describe accurately the capacity behavior for SNR values as small as $4\dB$. 

It is appropriate to point out that there is no guarantee that our upper bound $U(\snr)$ converges to the asymptotic capacity expression~\eqref{eq:capacity_asymptotic} as SNR grow large. In fact, the output distribution used in the duality step in the two cases is different. Obtaining a tighter non-asymptotic bound based on the output distribution that is optimal asymptotically remains an open problem.


Plots for the case of a non-unitary matrix $\matH$ can be obtained directly from Fig.~\ref{fig:6degMIMO} by shifting the upper bound to the left by  $\lambda\sub{max}$ (expressed in $\dB$), and shifting the lower bound to the right by  $\lambda\sub{min}$ (expressed in $\dB$).
The gap between the resulting upper and lower bounds increases proportionally to the logarithm of the ratio between $\lambda\sub{max}$ and   $\lambda\sub{min}$ in accordance to~\eqref{eq:upper_lower_gap}.

\section{Conclusions} 
\label{sec:conclusions}
We presented an asymptotic (high-SNR) characterization, as well as nonasymptotic bounds, on the capacity of MIMO microwave backhaul links affected by Wiener phase noise.
Our results are developed for the case of common oscillator at the transceivers, and under the practically relevant assumption that the transmit signal is subject to a peak-power constraint.
By numerical simulations, we showed that our asymptotic capacity expression, which---differently from the capacity upper and lower bounds---is trivial to compute, is accurate at the SNR values typically encountered in microwave backhaul links ($15\dB$ or higher).
In the regime where our asymptotic capacity formula is tight,  QAM constellations exhibit a gap of about $3$\dB. 
The gap to the capacity upper bound may be reduced by replacing QAM with suitably optimized constellations.
Furthermore, the upper bound could be further tightened by substituting the Gamma distribution~\eqref{eq:output_distribution_avp} used in the duality step with the asymptotically optimal output distribution~\eqref{eq:optimal_output_asympt}, and then by optimizing over the parameter~$\epsilon$.
This, however, may further increase the computational complexity associated to the numerical evaluation of the upper bound.



\begin{appendix}
\subsection{Proof of \fref{thm:large_snr}}
\label{sec:appendix}

  The asymptotic characterization~\eqref{eq:capacity_asymptotic} is obtained by proving that the upper bound~\eqref{eq:memoryless_plus_correction} matches up to a $\landauo(1)$ term the lower bound we shall report in Appendix~\ref{sec:lower_bound} below.
\subsubsection{Lower bound} 
\label{sec:lower_bound}
We take $\{\vecx_k\}$ \iid and isotropically distributed according to Proposition~\ref{propo:isotropic_distribution}.
Specifically, we let $\vecx_k\distas\sqrt{\snr}\vecz_{\snr,\snr_0}$ where for a given $\snr_0>0$ the random variable $\vecz_{\snr,\snr_0}$ has the following pdf:
\begin{IEEEeqnarray}{rCL}\label{eq:input_distribution}
  f_{\vecz_{\snr,\snr_0}}(\veca)=\frac{f_{\vecz}(\veca)}{\Pr\{\vecnorm{\vecz}^2\geq \snr_0/\snr\}}\indfun\lefto\{ \frac{\snr_0}{\snr}\leq\vecnorm{\veca}^2\leq 1
  \right\}. \IEEEeqnarraynumspace
\end{IEEEeqnarray}
Here, $\indfun\{\cdot\}$ denotes the indicator function
and the pdf $f_{\vecz}$ of the random variable $\vecz$ is given by
\begin{IEEEeqnarray}{rCL}\label{eq:pdf_z}
  f_{\vecz}(\veca)&=&\left(M-\frac{1}{2}\right) \frac{\Gamma(M)}{\pi^M}\frac{1}{\vecnorm{\veca}}\indfun\lefto\{\vecnorm{\veca}^2\leq 1\right\}.
\end{IEEEeqnarray}
Note that for every $\snr_0>0$ the pdf $f_{\vecz_{\snr,\snr_0}}$ converges pointwise to  $f_{\vecz}$ as $\snr\to\infty$.
The rationale behind the choice of $f_{\vecz}$ is that it turns out to maximize $h(\vecz)-\Ex{}{\log\vecnorm{\vecz}}$ under the constraint that $\vecnorm{\vecz}\leq 1$ \wpone. 
The pdf $f_{\vecz_{\snr,\snr_0}}$ is constructed from $f_\vecz$ so as to guarantee that  $\vecnorm{\vecx_k}^2\geq \snr_0$, a property that will be useful in the remainder of the proof.

To obtain the desired lower bound, we first proceed as in~\cite{durisi13-02a} and
use chain rule for mutual information and that mutual information is nonnegative to obtain
\begin{IEEEeqnarray}{rCL}
  I(\vecx^n;\vecy^n) &= &\sum_{k=1}^n I(\vecx_k; \vecy^n \given \vecx^{k-1}) \\
  & \geq &\sum_{k=2}^{n} I(\vecx_k; \vecy^k \given \vecx^{k-1}).\label{eq:chain_rule_step_lower_bound}
\end{IEEEeqnarray}
%
Fix now $k\geq 2$  and set
\begin{IEEEeqnarray*}{rCL}
  \epsilon_k\define I(\vecx_k;\theta_{k-1} \given \vecy_k,\vecy_{k-1},\vecx_{k-1}).
\end{IEEEeqnarray*}
We have
\begin{IEEEeqnarray}{rCL}
  I(\vecx_k;\vecy^k\given \vecx^{k-1}) &=& I(\vecx_k;\vecy^k, \vecx^{k-1})\label{eq:the_two_terms_lb_a}\\
  &{\geq}& I(\vecx_k;\vecy_k, \vecy_{k-1}, \vecx_{k-1})\label{eq:the_two_terms_lb_b}\\
  &=& I(\vecx_k;\vecy_k, \vecy_{k-1}, \vecx_{k-1}, \theta_{k-1}) -\epsilon_k \IEEEeqnarraynumspace\\
  &{=}& I(\vecx_k;\vecy_k, \theta_{k-1}) -\epsilon_{k}\label{eq:the_two_terms_lb_c}\\
  &{=}&I(\vecx_k;\vecy_k \given \theta_{k-1})-\epsilon_{k}\label{eq:the_two_terms_lb_d}\\
  &{=}& I(\vecx_{2};\vecy_{2} \given \theta_{1}) -\epsilon_{2}.\label{eq:the_two_terms_lb}
\end{IEEEeqnarray}
Here,~\eqref{eq:the_two_terms_lb_a} follows because the $\{\vecx_k\}$ are independent; in~\eqref{eq:the_two_terms_lb_b} we used chain rule for mutual information and that mutual information is nonnegative;~\eqref{eq:the_two_terms_lb_c} follows because $\vecx_k$ and the pair $(\vecy_{k-1},\vecx_{k-1})$ are conditionally independent given $(\theta_{k-1},\vecy_k)$;~\eqref{eq:the_two_terms_lb_d} holds because $\vecx_k$ and $\theta_{k-1}$ are independent; finally,~\eqref{eq:the_two_terms_lb} follows from stationarity.
Substituting~\eqref{eq:the_two_terms_lb} into~\eqref{eq:chain_rule_step_lower_bound} and then~\eqref{eq:chain_rule_step_lower_bound} into \eqref{eq:capacity}, we obtain
\begin{IEEEeqnarray}{rCL}\label{eq:lower_bound_with_two_terms}
  C(\snr)&\geq&  I(\vecx_{2};\vecy_{2} \given \theta_{1}) -\epsilon_{2}.
\end{IEEEeqnarray}
We next investigate the two terms on the RHS of~\eqref{eq:lower_bound_with_two_terms} separately.
We shall show that the first term has the desired asymptotic expansion, while the second term can be made arbitrarily close to zero by choosing $\snr_0$  sufficiently large.
%
\paragraph*{The first term on the RHS of~\eqref{eq:lower_bound_with_two_terms}} 
\label{par:the_first_term_on_the_rhs_of_eq:lower_bound_with_two_terms}
We write
\begin{align}\label{eq:lb_standard_decomposition_mi}
  I(\vecx_{2};\vecy_{2} \given \theta_{1})=\difent(\vecy_{2} \given \theta_1) - \difent(\vecy_{2} \given \vecx_{2},\theta_{1})
\end{align}
and bound the two terms separately.
For the first term, we have that
\begin{IEEEeqnarray}{rCL}
  \difent(\vecy_{2}\given \theta_1)  
  &\geq& \difent(\vecy_{2} \given \vecw_{2},\theta_1) \\
  &=&\difent(e^{j\theta_2}\vecx_{2}\given \theta_1)\\
  &=&\difent(\vecx_2)\label{eq:kill_noise_and_use_isotropy}\\
  &=& M\log \snr + \difent(\vecz_{\snr,\snr_0})\\
  &=& M\log \snr - \log\frac{(M-1/2)\Gamma(M)}{\pi^M \Pr\{\vecnorm{\vecz}^2\geq \snr_0/\snr\}} \notag \\
  &&+\ \Ex{}{\log\vecnorm{\vecz_{\snr,\snr_0}}}.\IEEEeqnarraynumspace\label{eq:entropy_of_input}
\end{IEEEeqnarray}
Here~\eqref{eq:kill_noise_and_use_isotropy} follows because $\vecx_2$ is isotropically distributed and~\eqref{eq:entropy_of_input} holds because of~\eqref{eq:input_distribution} and~\eqref{eq:pdf_z}.
For the second term on the RHS of~\eqref{eq:lb_standard_decomposition_mi}, we proceed as follows.
Let $\vecx_2=s_2\vecv_2$, with $s_2=\vecnorm{\vecx_2}$ and, hence,  $s_2^2\distas\snr\vecnorm{\vecz_{\snr,\snr_0}}^2$.
Furthermore, let $z_2\distas \jpg(0,1)$.
Then, proceeding as in~\cite[Eq.~(10)]{durisi12-08a}
\begin{IEEEeqnarray}{rCL}
  \difent(\vecy_2 \given \vecx_2,\theta_1)&=& \difent(\vecy_2\given s_2,\vecv_2,\theta_1) \\
  &=& \difent(e^{j\theta_2}s_2+z_2 \given s_2, \theta_1) +\log (\pi e)^{M-1}.\IEEEeqnarraynumspace\label{eq:cond_dif_ent_as_lb}
\end{IEEEeqnarray}
The first term on the RHS of~\eqref{eq:cond_dif_ent_as_lb} can be bounded as follows
\begin{IEEEeqnarray}{rCL}
  \IEEEeqnarraymulticol{3}{l}{ \difent(e^{j\theta_2}s_2+z_2 \given s_2, \theta_1)}\\
   %
   &=& \difent(e^{j\theta_2}(s_2+z_2) \given s_2, \theta_1)\\
   &{=}&\difent(e^{j\Delta}(s_2+z_2) \given s_2) \label{eq:cond_diff_ent_lb_a}\\ 
  &{=}& \difent(\abs{s_2+z_2}^2\givenalt s_2) \label{eq:cond_diff_ent_lb_b}\notag\\
  &&+ \difent(\phi_2(s^2_2)+\Delta\givenalt \abs{s_2+z_2}, s_2) -\log 2 \\
  &{\leq}& \frac{1}{2}\Ex{}{\log\lefto(2\pi e\left[1+2\snr\vecnorm{\vecz_{\snr,\snr_0}}^2\right]\right)} \label{eq:cond_diff_ent_lb_c}\notag\\
  && + \difent(\phi_2(s^2_2) + \Delta\given  s_2) -\log 2.
\end{IEEEeqnarray}
Here, in~\eqref{eq:cond_diff_ent_lb_a} we used~\eqref{eq:wiener_model} and denoted by $\Delta$ a random variable distributed as in~\eqref{eq:pdf_wiener_increment};
in~\eqref{eq:cond_diff_ent_lb_b} we evaluated the differential entropy in polar coordinates using~\cite[Lemma~6.15 and Lemma~6.16]{lapidoth03-10a}.
Finally,~\eqref{eq:cond_diff_ent_lb_c} follows because the Gaussian distribution maximizes differential entropy under a variance constraint and because conditioning reduces entropy.
Note that
\begin{IEEEeqnarray}{rCL}
  \difent(\phi_2(s_2^2)+\Delta\given  s_2)&\leq& \max_{\xi\geq \sqrt{\snr_0} } \difent(\phi_2(\xi^2)+\Delta) \notag\\
  &=&\difent(\phi_2(\snr_0)+\Delta).
\end{IEEEeqnarray}
This term can be made arbitrarily close to $h(\Delta)$ by choosing $\snr_0$ in~\eqref{eq:input_distribution} sufficiently large.
Summarizing, we have shown that
\begin{IEEEeqnarray}{rCL}
  \IEEEeqnarraymulticol{3}{l}{ I(\vecx_2;\vecy_2\given \theta_1)\geq M\log \snr - \log\frac{(M-1/2)\Gamma(M)}{\pi^M \Pr\{\vecnorm{\vecz}^2\geq \snr_0/\snr\}}} \notag\\
  &&+\ \Ex{}{\log\vecnorm{\vecz_{\snr,\snr_0}}} 
   - \frac{1}{2}\Ex{}{\log\lefto(2\pi e\left[1+2\snr\vecnorm{\vecz_{\snr,\snr_0}}^2\right]\right)}\notag\\
  && - \difent(\phi_2(\snr_0)+\Delta) +\log 2 -\log(\pi e)^{M-1}\\
  &=&  \left(M-\frac{1}{2}\right)\log\snr - \log\lefto(M-\frac{1}{2}\right) -\log \Gamma(M)\notag\\
&&+\frac{1}{2}\log \pi-\left(M-\frac{1}{2}\right)-\difent(\phi_2(\snr_0)+\Delta)+\landauo(1). \IEEEeqnarraynumspace
\end{IEEEeqnarray}
Here, the last step follows because
\begin{IEEEeqnarray}{rCL}
  \Ex{}{\log(1+c\snr\vecnorm{\vecz_{\snr,\snr_0}}^2)}&=&\log(c\snr)\notag\\
  &&+\Ex{}{\log\vecnorm{\vecz_{\snr,\snr_0}}^2}+\landauo(1)  \IEEEeqnarraynumspace
\end{IEEEeqnarray}
for all $c>0$, and
\begin{IEEEeqnarray}{rCL}
  \lim_{\snr\to \infty}\Pr\{\vecnorm{\vecz}^2\geq \snr_0/\snr\}=1.
\end{IEEEeqnarray}
%


\paragraph*{The second term on the RHS of~\eqref{eq:lower_bound_with_two_terms}} 
\label{par:the_second_term_on_the_rhs_of_eq:lower_bound_with_two_terms}
Let $\vecx_1=s_1\vecv_1$ and $z_1\distas\jpg(0,1)$. 
Proceeding similarly as in~\cite[App.~IX]{lapidoth03-10a}, we obtain (see~\cite[Eq.~(25)]{durisi13-02a})
\begin{IEEEeqnarray}{rCL}
  \IEEEeqnarraymulticol{3}{l}{I(\vecx_2;\theta_1\given \vecy_2,\vecy_1,\vecx_1)}\notag\\
  &=&\difent(\theta_2\given e^{j\theta_1}(\sqrt{\snr_0}+z_1))-\difent(\theta_2\given \theta_1)\label{eq:bound_on_epsilon}.
\end{IEEEeqnarray}
As claimed, the RHS of~\eqref{eq:bound_on_epsilon} can be made arbitrarily close to zero by choosing $\snr_0$ in~\eqref{eq:input_distribution} sufficiently large.


\subsubsection{Upper Bound} 
\label{sec:upper_bound}
We exploit the property that the high-SNR behavior of $C(\snr)$ does not change if the support of the input distribution is constrained to lie outside a sphere of arbitrary radius. 
This result, known as \emph{escape-to-infinity} property of the capacity-achieving input distribution~\cite[Def.~4.11]{lapidoth03-10a}, is formalized in the following lemma.
\begin{lem}\label{lem:escape-to-infty}
   Fix an arbitrary $\xi_0>0$ and let $\setK(\xi_0)=\{\vecx \in \complexset^M\sothat \vecnorm{\vecx}\geq \xi_0\}$.
   Denote by $C^{(\xi_0)}(\snr)$ the capacity of the channel~\eqref{eq:io} when the input signal is subject to the peak-power constraint~\eqref{eq:pap} and to the additional constraint that $\vecx_k \in \setK(\xi_0)$ almost surely for all $k$.  
Then
\begin{align}\label{eq:escape_to_infty}
	\capacity(\snr) = \capacity^{(\xi_0)}(\snr) + \landauo(1), \quad \snr \to \infty
\end{align}
with $\capacity(\snr)$ given in~\eqref{eq:capacity}.
\end{lem}
\begin{IEEEproof}
The lemma follows directly from~\cite[Th.~8]{lapidoth06-02a} and~\cite[Th.~4.12]{lapidoth03-10a}.
\end{IEEEproof}

Fix $\xi_0>0$. 
By proceeding as in~\eqref{eq:memoryless_plus_correction}, we obtain\footnote{To keep notation compact, we write $\vecx_0$ simply as $\vecx$; same convention for $\vecy_0$.} 
\begin{IEEEeqnarray}{rCL}\label{eq:ub_asymptotic}
  \capacity^{(\xi_0)}(\snr) \leq \sup \bigl\{ I(\vecy;\vecx) \bigr\} + \log(2\pi)- \difent(\Delta)
\end{IEEEeqnarray}
where, this time, the supremum is over all probability distributions on $\vecx$ that satisfy $\vecnorm{\vecx}^2 \in [\xi_0^2, \snr]$ \wpone.
We next upper-bound $I(\vecy;\vecx)$ by using duality as in~\eqref{eq:ub_bound_on_the_memoryless_part}, i.e., we exploit that
\begin{IEEEeqnarray}{rCL}\label{eq:standard_decomposition_mi}
  I(\vecy;\vecx)\leq -\Ex{}{\log q_{\vecy}(\vecy)}-\difent(\vecy\given \vecx)
\end{IEEEeqnarray}
for every output distribution $q_{\vecy}(\vecy)$.
We choose a different $q_{\vecy}(\vecy)$ than the one resulting in~\eqref{eq:output_distribution_avp}.
Roughly speaking, we want $q_{\vecy}(\vecy)$ to be the output distribution induced by the input distribution~\eqref{eq:input_distribution} we used for the lower bound.
When constructing $q_{\vecy}(\vecy)$, we shall ignore the additive noise  over the support of the input distribution, and consider the effect of the additive noise only outside an $\epsilon$-neighborhood of the set $\{\vecx \in \complexset^M \sothat \vecnorm{\vecx}^2\leq \snr\}$.\footnote{This choice is inspired by~\cite{koch12-07a}, where the rates achievable with dense constellations over an AWGN channel (no phase noise) are analyzed.}
Specifically, we shall set $\vecr\define\vecy/\sqrt{\snr}$, $\setS_{\epsilon}\define\{\vecr \in \complexset^M \sothat \vecnorm{\vecr-\vecx'}\leq \epsilon \text{ for some } \vecnorm{\vecx'}\leq 1\}$  and choose the following probability distribution for $\vecr$
\begin{IEEEeqnarray}{rCL}\label{eq:optimal_output_asympt}
  q_{\vecr}(\vecr)=\begin{cases}
   \displaystyle
  \frac{(M-1/2)}{\pi^M K_{\snr,\epsilon}}\frac{\Gamma(M)}{\vecnorm{\vecr}}, \quad \text{if} \quad \vecr \in \setS_{\epsilon}\\[4mm]
  \displaystyle
  \frac{\snr^M}{\pi^M K_{\snr,\epsilon}} e^{-\snr\vecnorm{\vecr}^2}, \quad \text{if} \quad \vecr \notin \setS_{\epsilon}
  \end{cases}
\end{IEEEeqnarray}
where 
\begin{IEEEeqnarray}{rCL}\label{eq:norm_const}
  K_{\snr,\epsilon} &=&\underbrace{\int_{\vecr \in \setS_{\epsilon}} \frac{(M-1/2)}{\pi^M}\frac{\Gamma(M)}{\vecnorm{\vecr}} d\vecr}_{\define K_{\infty,\epsilon}} \notag \\
  &&+ \int_{\vecr \notin \setS_{\epsilon}} \frac{\snr^M}{\pi^M } e^{-\snr\vecnorm{\vecr}^2} d\vecr.
\end{IEEEeqnarray}
This yields 
\begin{IEEEeqnarray}{rCL}\label{eq:duality_with_vecr}
  -\Ex{}{\log q_{\vecy}(\vecy)}&=& M \log \snr -\Ex{}{\log q_{\vecr}(\vecr)}
\end{IEEEeqnarray}
where
\begin{IEEEeqnarray}{rCL}
  -\Ex{}{\log q_{\vecr}(\vecr)} &=& -\left[\log\frac{(M-1/2)\Gamma(M)}{\pi^M K_{\snr,\epsilon}} \right]\Pr\{\vecr \in \setS_{\epsilon}\}\notag\\
  &&+\Ex{}{(\log\vecnorm{\vecr})\indfun\{\vecr \in \setS_{\epsilon}\}} \notag \\
  &&-\left[\log\frac{\snr^M}{\pi^M K_{\snr,\epsilon}}\right] \Pr\{\vecr \notin\setS_{\epsilon}\}\notag\\
  && + \Ex{}{\snr\vecnorm{\vecr}^2 \indfun\{\vecr \notin \setS_{\epsilon}\}}.\label{eq:duality_step_with_vecr}
\end{IEEEeqnarray}
We next characterize each term on the RHS of~\eqref{eq:duality_step_with_vecr} in the limit $\snr \to \infty$.
\paragraph*{The first term} 
By construction (see~\eqref{eq:norm_const}), we have that 
\begin{IEEEeqnarray}{rCL}\label{eq:norm_constant_lim}
  \lim_{\epsilon \to 0}\lim_{\snr \to \infty} K_{\snr,\epsilon} =\lim_{\epsilon \to 0} K_{\infty,\epsilon}=1.
\end{IEEEeqnarray}
Furthermore, let $\vecw\distas\jpg(\veczero,\matI_M)$.
Then
\begin{IEEEeqnarray}{rCL}
  \Pr\{\vecr \notin \setS_{\epsilon}\}&=& \Pr\{\vecnorm{\vecw}/\sqrt{\snr}\geq \epsilon\}\label{eq:epsilon_neigh_trick}\\
  &=& \frac{\Gamma(M,\epsilon^2 \snr)}{\Gamma(M)}\label{eq:tail_chi_square}\\
  &=&\frac{(\epsilon^2\snr)^{M-1}e^{-\epsilon^2 \snr}}{\Gamma(M)} +\landauo(\snr^{M-1} e^{-\epsilon^2\snr})\label{eq:tail_chi_square_asymptotic}
\end{IEEEeqnarray}
for $\quad \snr\to\infty$. Here,~\eqref{eq:epsilon_neigh_trick} follows because $\vecr\distas \vecx'+\vecw/\sqrt{\snr}$ for some $\vecnorm{\vecx'}\leq 1$;
in~\eqref{eq:tail_chi_square}, the function $\Gamma(\cdot,\cdot)$ is the upper incomplete Gamma function~\cite[Eq.~6.5.3]{abramowitz72-a}; finally,~\eqref{eq:tail_chi_square_asymptotic} follows from~\cite[Eq.~6.5.32]{abramowitz72-a}.
Using~\eqref{eq:tail_chi_square_asymptotic}, we conclude that the first term on the RHS of~\eqref{eq:duality_step_with_vecr} admits the following asymptotic expansion: 
\begin{multline}\label{eq:first_term}
  \left[\log\frac{(M-1/2)\Gamma(M)}{\pi^M K_{\snr,\epsilon}} \right]\Pr\{\vecr \in \setS_{\epsilon}\}\\=\log\frac{(M-1/2)\Gamma(M)}{\pi^M K_{\infty,\epsilon}}+\landauo(1).
\end{multline}
Furthermore, \eqref{eq:norm_constant_lim} implies that $K_{\infty,\epsilon}$ can be made arbitrarily close to $1$ by choosing $\epsilon$ sufficiently small. 
\paragraph*{The second term} 
\label{par:the_second_term}
Note that
\begin{multline}
  \Ex{}{(\log\vecnorm{\vecr})\indfun\{\vecr \in \setS_{\epsilon}\}} \\
  = \frac{1}{2}\Ex{}{(\log \vecnorm{\vecy}^2)\indfun\{\vecy/\sqrt{\snr}\in \setS_{\epsilon}\}} -\frac{1}{2}\log \snr.\label{eq:second_term}
\end{multline}
Assume without loss of generality that $\snr>1$.
Then $(\vecy/\sqrt{\snr}) \notin \setS_{\epsilon}$ implies that $\vecnorm{\vecy}^2>\snr>1$.
Hence, we conclude that $\log \vecnorm{\vecy}^2>0$ whenever $(\vecy/\sqrt{\snr}) \notin \setS_{\epsilon}$.
As a consequence, we can upper-bound~\eqref{eq:second_term} by adding 
\begin{IEEEeqnarray}{rCL}
  \frac{1}{2}\Ex{}{(\log \vecnorm{\vecy}^2)\indfun\{\vecy/\sqrt{\snr}\notin \setS_{\epsilon}\}}
\end{IEEEeqnarray}
and obtain 
\begin{IEEEeqnarray}{rCL}
  \Ex{}{(\log\vecnorm{\vecr})\indfun\{\vecr \in \setS_{\epsilon}\}} &\leq& \frac{1}{2}\Ex{}{\log \vecnorm{\vecy}^2} -\frac{1}{2}\log \snr.
\end{IEEEeqnarray}
%


\paragraph*{The third term} 
\label{par:the_third_term}
It follows from~\eqref{eq:tail_chi_square_asymptotic} that
\begin{IEEEeqnarray}{rCL}
  \left[\log\frac{\snr^M}{\pi^M K_{\snr,\epsilon}}\right] \Pr\{\vecr \notin\setS_{\epsilon}\}=\landauo(1).\label{eq:third_term}
\end{IEEEeqnarray}
%


\paragraph*{The fourth term} 
\label{par:the_fourth_term}
We have that
\begin{IEEEeqnarray}{rCL}
  \Ex{}{\snr\vecnorm{\vecr}^2\indfun\{\vecr \notin \setS_{\epsilon}\}} &\leq& \snr \sqrt{\Ex{}{\vecnorm{\vecr}^4}\Pr\{\vecr \notin \setS_{\epsilon}\}}\label{eq:cauchy}\\
  &=&\landauo(1).\label{eq:fourth_term}
\end{IEEEeqnarray}
Here,~\eqref{eq:cauchy} follows from Chaucy-Schwarz inequality and~\eqref{eq:fourth_term} follows from~\eqref{eq:tail_chi_square_asymptotic}.

We next substitute~\eqref{eq:first_term},~\eqref{eq:second_term},~\eqref{eq:third_term}, and~\eqref{eq:fourth_term} into~\eqref{eq:duality_step_with_vecr} and then~\eqref{eq:duality_step_with_vecr} into~\eqref{eq:duality_with_vecr} and obtain
\begin{IEEEeqnarray}{rCL}\label{eq:bound_on_first_term}
  -\Ex{}{\log q_{\vecy}(\vecy)}&=&\left(M-\frac{1}{2}\right)\log \snr - \log \frac{(M-1/2)\Gamma(M)}{\pi^M K_{\infty,\epsilon}}\notag \\
  &&+\ \frac{1}{2}\Ex{}{\log \vecnorm{\vecy}^2} + \landauo(1).
\end{IEEEeqnarray}

Set now $s=\vecnorm{\vecx}$ and $z\distas \jpg(0,1)$.
By proceeding as in~\cite[Eq.~(33)]{durisi13-02a}, we can rewrite the conditional differential entropy $h(\vecy\given \vecx)$ on the RHS of~\eqref{eq:standard_decomposition_mi} as
\begin{IEEEeqnarray}{rCL}\label{eq:cond_diff_ent}
  \difent(\vecy\given \vecx)= \difent(\abs{s+z}^2 \given s) + \log \pi^M + M-1.
\end{IEEEeqnarray}
Substituting~\eqref{eq:bound_on_first_term} and~\eqref{eq:cond_diff_ent} into~\eqref{eq:standard_decomposition_mi}
and using that
\begin{IEEEeqnarray}{rCL}
  \vecnorm{\vecy}^2&\distas& \abs{s+z_1}^2 + \sum_{j=2}^M\abs{z_j}^2
\end{IEEEeqnarray}
where $z_j\distas \jpg(0,1)$, $j=1,\dots, M$, we obtain
\begin{IEEEeqnarray}{rCL}
  \IEEEeqnarraymulticol{3}{l}{I(\vecy;\vecx)\leq  \left(M-\frac{1}{2}\right)\log \snr - \log \frac{(M-1/2)\Gamma(M)}{K_{\infty,\epsilon}}}\notag\\
  %
  &&+\ \frac{1}{2}\Exop \Biggl[\log \Biggl(\abs{s+z_1}^2 + \sum_{j=2}^M\abs{z_j}^2\Biggr)\Biggr] \notag \\
  && -\ \difent(\abs{s+z}^2 \given s) -(M-1) +\landauo(1) \\
  &\leq& \left(M-\frac{1}{2}\right)\log \snr - \log \frac{(M-1/2)\Gamma(M)}{K_{\infty,\epsilon}}-(M-1)\notag\\ 
  &&+\max_{\xi_0\leq \xi\leq \sqrt{\snr}} \Biggl\{\frac{1}{2}\Exop \Biggl[\log \Biggl(\abs{\xi+z_1}^2 + \sum_{j=2}^M\abs{z_j}^2\Biggr)\Biggr] \notag \\
  && -\difent(\abs{\xi+z}^2)\Biggr\}+\landauo(1)\label{eq:final_bound}.
\end{IEEEeqnarray}

Substituting~\eqref{eq:final_bound} into~\eqref{eq:ub_asymptotic} and using that
\begin{multline}
  \lim_{\xi\to\infty} \Biggl\{\frac{1}{2}\Exop\Biggl[\log\Biggl(\abs{\xi+z_1}^2 + \sum_{j=2}^M\abs{z_j}^2\Biggr)\Biggr] - \difent\lefto(\abs{\xi+z}^2\right)\Biggr\}\\= -\frac{1}{2}\log(4\pi e)
\end{multline}
which follows by \cite[Eq.~(9)]{lapidoth02-10a} and by proceeding similarly to the proof of~\cite[Lemma~6.9]{lapidoth03-10a},
we conclude that we can make the bound on $\capacity^{(\xi_0)}(\snr)$ just derived to be arbitrarily close to~\eqref{eq:capacity_asymptotic} in the high-SNR regime by choosing $\epsilon$ sufficiently small and $\xi_0$ sufficiently large.

\end{appendix}


\end{document}